\documentclass[12pt,hidelinks]{article}
\usepackage[T1]{fontenc}
\usepackage[utf8]{inputenc}
\usepackage{graphicx}
\usepackage{amsmath}
\usepackage{bm}
\usepackage{amssymb}
\usepackage{xcolor}
\usepackage{soul}
\usepackage{sectsty}
\usepackage{tgtermes}
\usepackage[labelfont={bf},font={small},figurename=Figure]{caption}
\usepackage[a4paper, total={6in, 8in}]{geometry} 
\usepackage[style=nature,articletitle=true,backend=biber,sorting=none,subentry,maxbibnames=99]{biblatex}
\usepackage{filecontents}
\sectionfont{\large}
\usepackage{float}
\usepackage{wrapfig}
\usepackage{authblk} 

\title{Messy or Ordered? Multi-scale Mechanics Dictates Shape-Morphing of Hierarchical 2D Fiber-Networks}
\author[1,2]{Shiran Ziv Sharabani}
\author[1,2]{Nicole Edelstein-Pardo}
\author[1,2]{Maya Molco}
\author[1]{Michael Morami}
\author[1]{Netanel Bachar Schwartz}
\author[1]{Aya Sivan}
\author[3]{Eli Flaxer}
\author[1,2,$\ast$]{Amit Sitt}

\affil[1]{\normalsize{Department of Physical Chemistry, The School of Chemistry, Raymond and Beverly Sackler Faculty of Exact Sciences, Tel-Aviv University, Tel Aviv 6997801, Israel} }
\affil[2]{The Center for Nanoscience and Nanotechnology, Tel Aviv University, Tel Aviv 6997801, Israel}
\affil[3]{AFEKA - Tel-Aviv Academic College of Engineering, 69107 Tel-Aviv, Israel}
\affil[$\ast$]{To whom correspondence should be addressed; E-mail: amitsitt@tauex.tau.ac.il}
\date{}

\begin{document}
\maketitle
\begin{abstract}
   Shape-morphing networks of mesoscale filaments are a common hierarchical feature in biology and hold significant potential for a range of technological applications, from micro-muscles to shape-morphing optical devices. Here, we demonstrate both experimentally and computationally that the shape-morphing of highly-ordered 2D networks constructed of thermoresponsive mesoscale polymeric fibers strongly depends on the physical attributes of the single fiber, in particular on its diameter, as well as on the network's density. We show that based on these parameters, such fiber-networks exhibit one of two thermally driven morphing behaviors: (i) the fibers stay straight, and the network preserves its ordered morphology, exhibiting a bulk-like behavior; or (ii) the fibers buckle and the network becomes messy and highly disordered. Notably, in both cases, the networks display memory and regain their original ordered morphology upon shrinking. This hierarchical principle, demonstrated here on a range of networks, offers a new way for controlling the shape morphing of materials with mesoscale resolutions, and elucidates that minute changes in the mesoscale structural attributes can translate to a dramatic change in the morphing behaviors at the macroscale.
\end{abstract}
Polymer-based shape-morphing materials that can autonomously change their morphology upon exposure to a stimulus are increasingly deployed for a range of applications  \supercite{Liu2016a,Gracias2013,White2015,Luo2019} including soft robotics \supercite{Hines2017}, artificial muscles \supercite{Park2020}, dynamic optics \supercite{Sun2012}, and microfluidics \supercite{Dong2006,Dong2007,TerSchiphorst2018}.
To fully realize the potential of these systems, there has been an ongoing effort to explore the mechanisms that govern the stimuli-induced morphology changes and to develop new morphing systems whose shape-morphing can be programmed and controlled. 

Although shape-morphing has been mainly examined at the macroscale, in particular in shape-morphing 2D thin sheets \supercite{Klein2007,Stoychev2011,Zakharchenko2010,Wu2013,Bae2017,Zhang2011} and in 2D networks of macroscale filaments \supercite{Gladman2016,Boley2019}, forming morphing systems in the mesoscale, which range from hundreds of nanometers to a few micrometers, is highly challenging. The fabrication of mesoscale systems has been mostly achieved using photolithography \supercite{Chang2018,Fan2016,Liu2021} and recently by using multiphoton lithography \supercite{Kaehr2008,Sun2012,Lee2015a,Wei2017}, and in particular focused on the formation of hollow 3D structures by folding or rolling of mesoscale 2D cutouts \supercite{Leong2010,Kempaiah2014,Erb2013,Kelby2011,Ge2013}.

An appealing approach for obtaining morphing with mesoscale resolution is to assemble morphing mesoscale building blocks such as mesoscale filaments into an hierarchical architecture. For example, thermoresponsive nanofibers were electrospun into dense and disordered non-woven shape-morphing meshes that exhibited a much faster shape-morphing rate than bulk polymer sheets of similar dimensions \supercite{Meng2015,Jiang2015,Jalani2014,Moon2021,Apsite2017,Zhao2018}.
Although hierarchical structuring of non-responsive mesoscale filaments was shown to induce significantly different mechanical properties from the bulk \supercite{Meza2014,Moroni2006,Chang2021,Mueller2021}, the effect of such multiscale hierarchy on the morphing behavior and mechanical properties of shape-morphing structures has not been examined to date.

Here, we report the construction of highly ordered 2D networks of interconnected thermoresponsive mesoscale fibers made of a poly-N-isopropylacrylamide (PNIPAAm) derivative, with high control over the fiber diameter and the fiber density. We demonstrate that both factors have a dramatic effect on the morphing behavior at the mesoscale level, which translates to the entire network, determining whether it would exhibit significant swelling and keep its ordered shape upon swelling, or if it would become messy and exhibit negligible swelling. The effect of these parameters on the mechanical properties of the filaments and on the morphing of the network is analyzed using a simple numerical simulation, which enables predicting the morphing behavior of the network based on the mechanical properties of the bulk polymer in its swollen and unswollen state.

\section*{Results}
\begin{figure}[H]
    \centering
    \includegraphics[width=\textwidth]{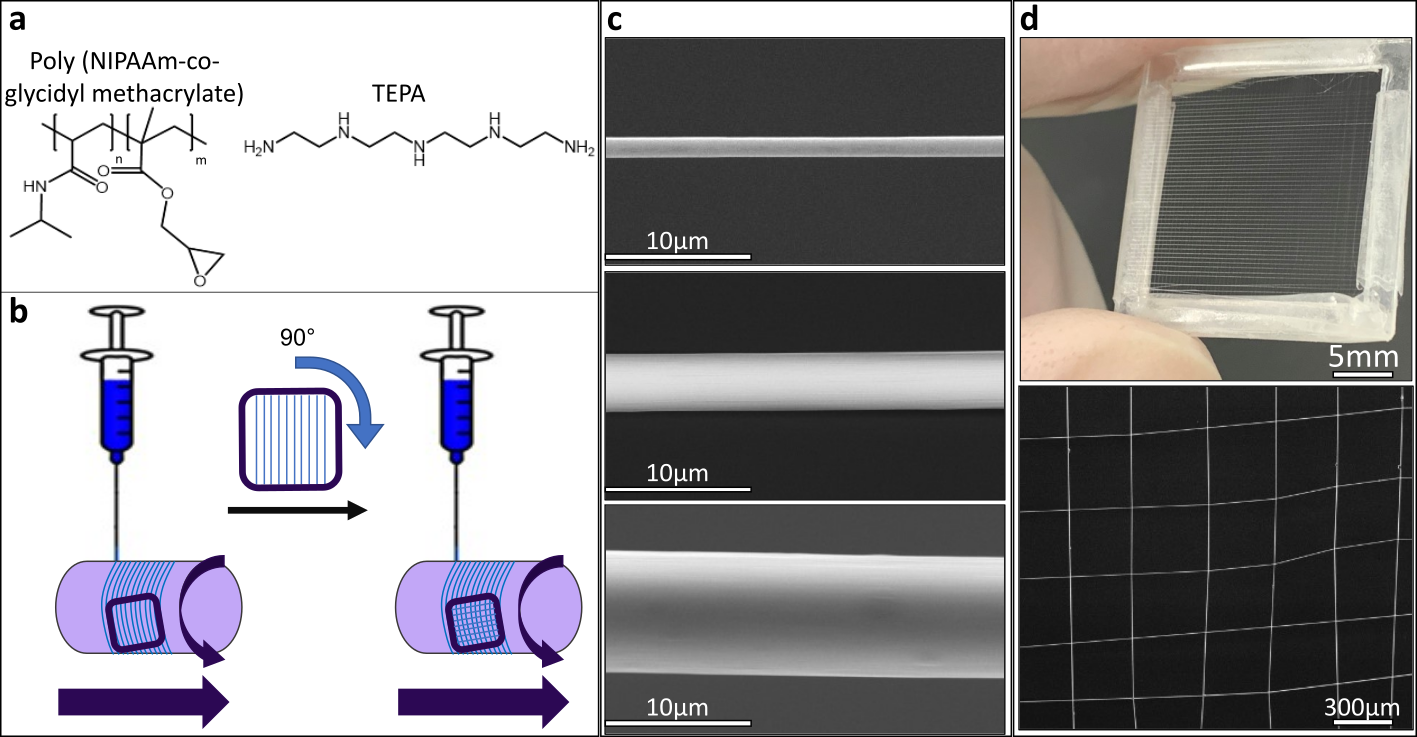}
    \caption{\textbf{(a)} The chemical structures of poly (NIPAAm-co-GMA) (left), and the crosslinker TEPA (right). \textbf{(b)} Schematic illustration of the dry-spinning process of the polymeric networks. \textbf{(c)} Scanning electron microscope (SEM) micrographs of as-spun microfibers with a diameter of 1.40 $\mu m$ (top), a diameter of 3.75 $\mu m$ (middle), and (bottom)  a diameter of 10.8 $\mu m$ (middle). \textbf{(d)} A network with fiber diameter 4.0 $\mu m$ and fiber density of 3 $mm^{-1}$ spun over a plastic frame (top), and its SEM micrograph (bottom).} 
    \label{fig:as_spun}
\end{figure}

The responsive polymer that was chosen for the fabrication of the networks is poly (N- isopropylacrylamide -co-glycidyl methacrylate) (PNcG) (monomers ratio of 50:1), a thermoresponsive derivative of the well-studied stimuli-responsive PNIPAAm. The PNcG copolymer was chosen because it can be crosslinked post-synthesis via the glycidyl groups by using multi-arm amine linkers (Fig. \ref{fig:as_spun}a). In an aqueous environment under ambient conditions, the crosslinked copolymer absorbs water and swells significantly to form a hydrogel. However, upon heating above its lower critical solution temperature (LCST) of 32$^{\circ}$C, the crosslinked copolymer expels the water and collapses, significantly reducing the gel volume  \supercite{Halperin2015,Schild1992}. 

2D networks were obtained by using the dry-spinning method (Fig. \ref{fig:as_spun}b). In this approach, the polymer and crosslinker were dissolved in appropriate solvents, and the solution was dispensed via a capillary. The fiber was drawn from the tip of the capillary and collected over a rotating frame that was placed on a moving stage, which allowed arranging the fibers in parallel with well-defined distances. The diameter of the fiber was controlled via the rotation speed (Fig. \ref{fig:as_spun}c).

Next, the frame was turned 90 degrees and the process was repeated to obtain the network 

\noindent (Fig. \ref{fig:as_spun}d). At the final stage, the networks were thermally crosslinked in an oven. 

Once crosslinked, the networks were cut from the frame and immersed in an aqueous environment on a heating glass slide that allows the system's temperature to be controlled. The shape-morphing process and the morphology of the networks upon heating and cooling were examined in both bright field and fluorescence modes using a light microscope. When the networks were immersed in water at a temperature above 32$^{\circ}$C, they kept their as-spun Cartesian morphology. Despite the thinness of the fibers and regardless of the fiber density, the networks also kept their planar 2D morphology, and did not exhibit buckling or folding, indicating the resilience of the 2D network architecture. In all the systems examined here, upon cooling below 32$^{\circ}$C to room temperature, the copolymer swelled significantly, as clearly indicated by the significant elongation in the fibers' contour length. The response of the fibers was extremely rapid and hence the morphing was governed mostly by the heating/cooling rate, which was much slower than the morphing rate of the fibers.

\begin{figure}[H]
    \centering
    \includegraphics[width=\textwidth]{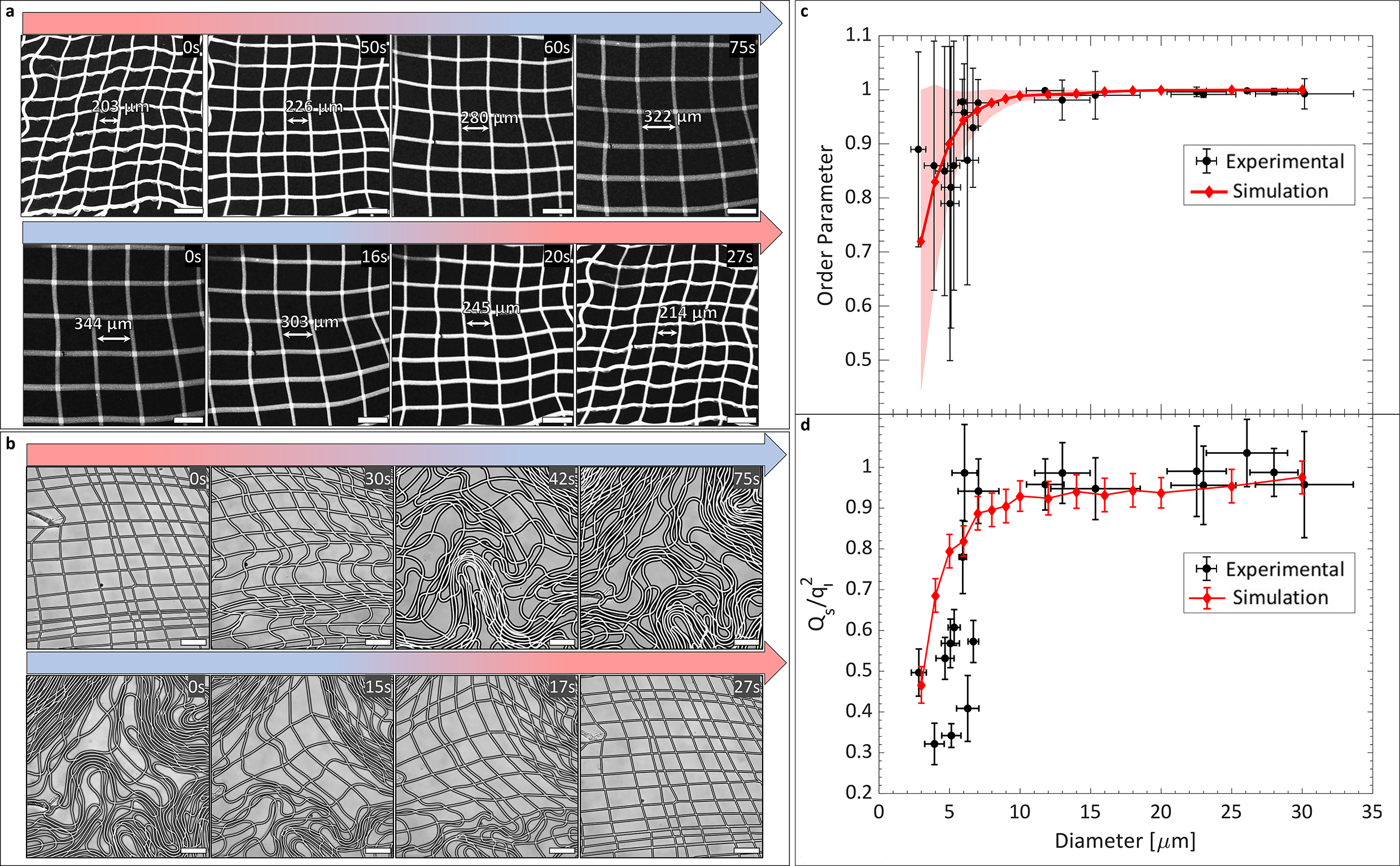}
    \caption{\textbf{(a)} Time lapse of the morphing of a shape-preserving network (Table S1, sample 16) taken while cooling from 60$^{\circ} C$ to 25$^{\circ}C$ (top) and while heating from 25$^{\circ}C$ to 60$^{\circ}C$ (bottom). Scale bar is $300 \mu m$. The average fiber diameter above LCST is $23.0 \mu m$ (SD $2.35 \mu m$). The average swelling ratio of the contour length is $1.70$ (SD $0.15$). \textbf{(b)} Time lapse depicting the morphing of a buckling-governed network (Table S1, sample2) while cooling from 60$^{\circ} C$ to 25$^{\circ} C$ (top), and while heating from 25$^{\circ} C$ to 60$^{\circ} C$ (bottom). Scale bar is $100 \mu m$. The average fiber diameter above the LCST is $3.93 \mu m$ (SD $0.69 \mu m$). The average swelling ratio of the contour length is $1.99$ (SD 0.30). \textbf{(c)} The OP of different networks (black) as a function of average fiber diameter. Error bars represent the standard deviations. The simulated OP obtained from the model is marked with a solid red line (shaded area represents the SD). \textbf{(d)} The ratio $Q_{S} / q_l^2$ of different networks as function of average fiber diameter (black). Error bars represent the standard deviations. The simulated ratio calculated from the model is marked with a solid line.}
    \label{fig:morph}
\end{figure}

The response of the networks to the temperature change could be divided into two distinct types based and their morphing behavior upon the swelling of the fibers. In the first type of networks, denoted as shape-preserving networks, upon swelling the fibers remained straight and the entire networks swelled and expanded homogeneously in all directions, preserving the original Cartesian morphology (Fig. \ref{fig:morph}a, top and movie S1). Upon reheating the system above 32$^{\circ}$C, the fibers went through a shrinking process, maintained their morphology, and preserved their shape (Fig. \ref{fig:morph}a, bottom). Such behavior resembles a bulk-like swelling, in which the swelling ratio of a small segment is similar to the swelling ratio of the entire system. 

In the second type of network, upon cooling the system below 32$^{\circ}$C, the extension of each fiber upon swelling resulted in the build-up of elastic instabilities, and the fibers exhibited substantial in-plane buckling into the voids of the network. Consequently, the final morphology of the swollen network did not keep its initial Cartesian morphology and the network became ”messy” and disordered (Fig. \ref{fig:morph}b, top panel, and movie S2). When heated above 32$^{\circ}$C, the fibers underwent fast-shrinking. The shrinking resulted in stretching and straightening of each fiber and the system recovered its original Cartesian form (Fig. \ref{fig:morph}b, bottom panel). Therefore, despite the evident disorder in the swollen form, the networks possessed shape-memory in their deflated form. Unlike in the case of the shape-preserving networks, the elongation of the fibers did not lead to the translation of the nodes but mainly to buckling of the fibers between the nodes of the network. Consequently, the change in the total area of the network in the swelling was much
\noindent lower than in the bulk or in shape-preserving networks. 

As the main change in the network swelling behavior is reflected in the order of the network, we constructed an order parameter (OP) that can quantify the order and allows distinguishing between the two types of shape-morphing behaviors. Inspired by OP in magnetic systems, each edge that connects two nodes in the network is represented as a chain of $N$ consecutive vectors of similar length, $\bm{v_i}$. The OP of the edge ,$s$, is defined as: 
\begin{equation}
    s=\frac{|\sum_{i=1}^n \bm{v}_i|}{N}.
    \label{op}
\end{equation}
For a straight edge, such as obtained in the as-spun network and in the swollen state of a shape-preserving network, $s$ is 1. The OP of the entire network is defined as the average of $s$ for all the edges constructing the network. 

Using this definition, the effect of changing different parameters on the morphing behavior of networks was investigated. We first examined the effect of the diameter of the fibers on their morphing behavior sparse networks. The OP for different networks as a function of their average fiber diameter (black) are shown in Fig. \ref{fig:morph}C. Error bars indicate the standard deviation. All the networks that were constructed of fibers with a diameter of 10 $\mu m$ and above (up to a fiber diameter of 30 $\mu m$) exhibited an OP of $\sim$1, suggesting that the fibers remain straight and induce a shape-preserving morphing mode. As the diameter of the fibers decreases from 10 to 5 $\mu m$, a gradual decrease in the OP is observed, indicating that as the diameter decreases, upon swelling, the buckling of the fibers become prominent and the networks become less ordered and messier. Below a diameter of 5 $\mu m$, buckling of the fibers is dominant upon swelling. Consequently, the OP significantly decreases and the SD increases, indicating the formation of a messy configuration. The gradual change in OP indicates the existence of a size-effect that determines the morphing behavior of the individual mesoscale fibers and consequently the morphing behavior of the networks. 

This size-dependence is also apparent when comparing the increase in the projected area of the network, $Q_s=S_{swollen}/S_{unswollen}$, and the longitudinal extension of the fibers upon swelling, $q_{l}=l_{swollen}/l_{unswollen}$. In networks with a diameter above 10 $\mu m$, $Q_s/q_{l}^2=1$, which indicates that the entire elongation of the fibers translates to the extension of the network' projection area, as expected if the network only scales in the swelling process. However, as the fiber diameter decreases below 10 $\mu m$, the swelling of the entire network is significantly smaller than the longitudinal swelling of the fibers because the buckling reduces the contribution of the fiber elongation to the swelling of the network (Fig. \ref{fig:morph}d).

\begin{figure}[t]
    \centering
    \includegraphics[width=0.6\textwidth]{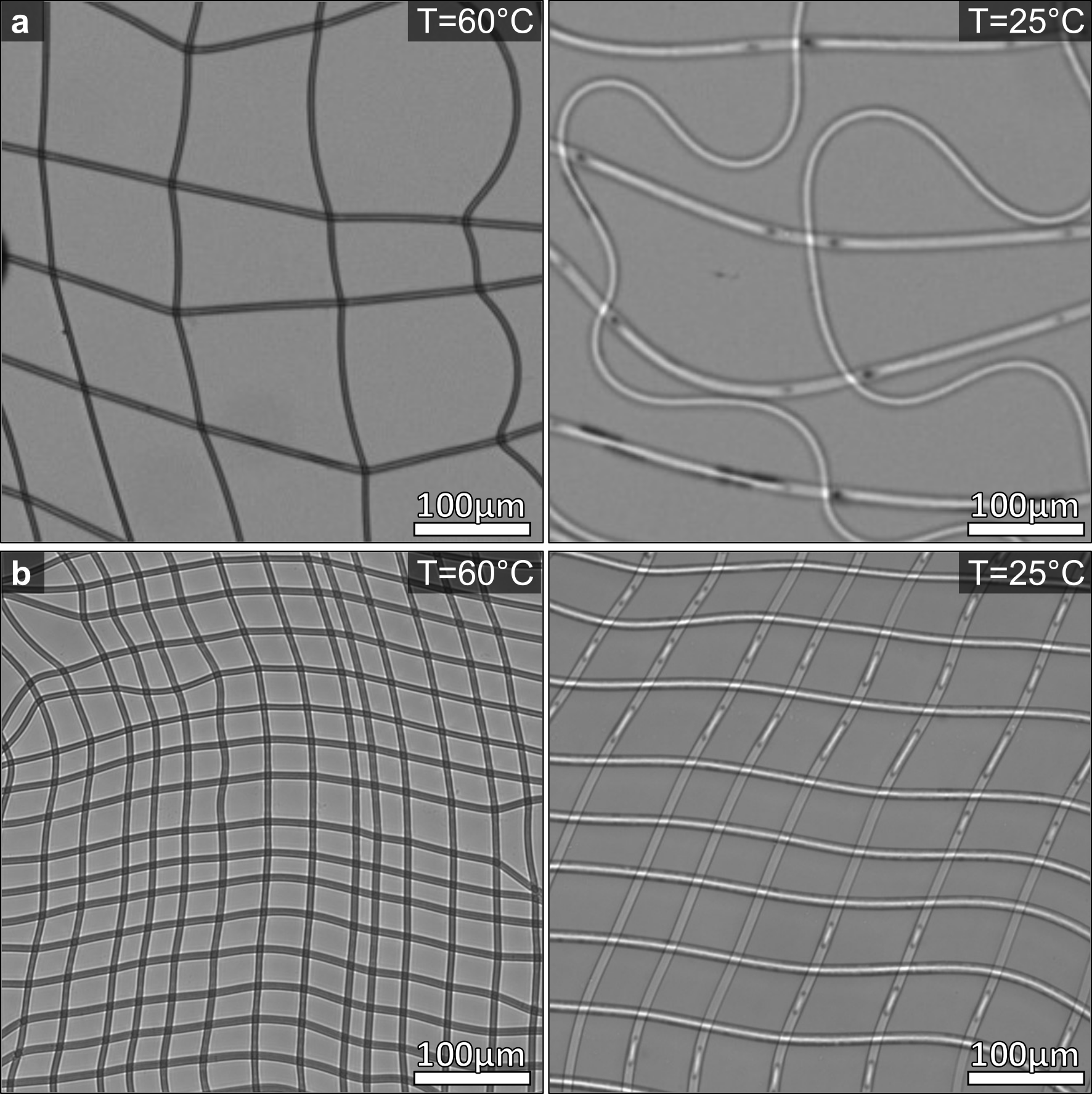}
    \caption{The effect of network density on the morphing behavior. \textbf{(a)} A sparse network of fiber density of $9mm^{-1}$ (SD $3mm^{-1}$) upon heating (left) and  cooling (right) exhibited a buckling-governed behavior with OP of $0.78$ (SD $0.21$). \textbf{(b)} A dense network with fiber density of $34mm^{-1}$ (SD $6\,mm^{-1}$) upon heating (left) and cooling (right) exhibited a shape-preserving behavior with OP of $0.984$ (SD $0.018$). Both networks have an average fiber diameter of $5.0\, \mu m$ (SD $1.1\, \mu m$) above the LCST.}
    \label{fig:dense}
\end{figure}

The second parameter that was examined is the fiber density of the network. As mentioned, for relatively thin fibers with a low density, the network is disordered upon cooling below the LCST. Two networks constructed of fibers of a similar averaged diameter that were fabricated using the same dry-spinning process are shown in Fig. \ref{fig:dense}. Fig. \ref{fig:dense}a shows a network with a low fiber density of $9\,mm^{-1}$ (SD $3\, mm^{-1}$) above the LCST, which exhibit a buckling-governed network upon cooling. Fig. \ref{fig:dense}b shows the behavior of a highly dense network with a density of $34\, mm^{-1}$ (SD $6\, mm^{-1}$) above the LCST. In the case of a high density, the network behaves as a shape-preserving network even though the fibers are relatively thin. This occurs when the distance between neighboring nodes is smaller than the characteristic buckling length of the fiber. Thus, even with thin fibers that tend to buckle, decreasing the length of the edge can be used to restrict its buckling and to induce a bulk-like swelling behavior.

To explain these trends in the morphing behavior, we regard the system as a network of interconnected slender beams that change their dimensions (diameter and length) and their Young modulus throughout the swelling process. For such a filament, the elastic energy is a sum of the stretching and bending energies:
\begin{equation}
    E_{elastic}(r,L_0,E)=\int_{0}^{L_0}{\frac{1}{2}k(r,L_0,E)\left(\frac{\partial l}{\partial l_0}-1\right)^2dl_0}+\int_{0}^{L_0}{\frac{1}{2}B(r,E)\left(\frac{\partial\phi}{\partial l_0}\right)^2dl_0},
        \label{energy}
\end{equation}
where $E$ is the Young modulus of the swollen polymer, $r$ is the radius of the filament, $L_0$ is the reference contour length of the filament before the deformation, $l_0$ and $l$ are the reference (rest length) and the deformed filament paths, $\phi$ is the local tangent angle, and $\frac{\partial \phi}{\partial l_0}=\frac{\partial ^2 l}{\partial l_0 ^2 }$ is the curvature of the filament. $k$ and $B$ are the stiffness and bending moduli, respectively, and are given by \supercite{elasticity}:
 
\begin{equation}
    k=E\frac{\pi r^2}{L_0},\quad B=E\frac{\pi r^4}{4}.
    \label{stiff}
\end{equation}
Another energy term is related to the spatial overlap between the fibers. While fibers can buckle out of the network plane, fiber overlap along the plane was rarely exhibited experimentally, indicating overlap is highly unfavorable. Thus, the fiber overlap energy term was regarded as a Heaviside step function. For $d \leq 2r$, where $d$ is the distance between two fibers the energy is high, and for $d > 2r$ the energy goes to zero.

Throughout the swelling process, significant loads are exerted on the edges constructing the network. If the load exerted on the edges is larger than the critical Euler buckling load given by:
\begin{equation}
    \sigma_{critical}=\frac{\pi^2 E r^4}{L_0^2},
    \label{E_overlap}
\end{equation}
the edges are expected to exhibit buckling instabilities and to diverge from the equilibrium morphology. The critical load depends on the fiber radius to the fourth power, which induces a strong size dependency. As the diameter of the fibers increases, the critical load rapidly increases as well and buckling is expected to be suppressed. This leads to a stretching-governed morphing and to shape-preservation of the network structure upon swelling. However, as the diameter of the fibers decreases, the critical buckling load becomes smaller, and thus, for thinner fibers buckling  becomes accessible and is expected to lead to a buckling-governed morphing. While these equations are general for a fiber under a load, for fibers constructed of stimuli-responsive polymers the dimensions of the fibers, $L_0$ and $r$, as well as the Young modulus change continuously throughout the swelling process, and these changes should be taken into account when describing their mechanical behavior. Furthermore, while in bulk 2D systems of non-compressible shape-morphing polymers the entire morphing information is manifested in the metric dictated by the local swelling, this is not necessarily the case for the networks in which buckling can occur and hence can be regarded as compressible.

\begin{figure}[t]
    \centering
    \includegraphics[width=\textwidth]{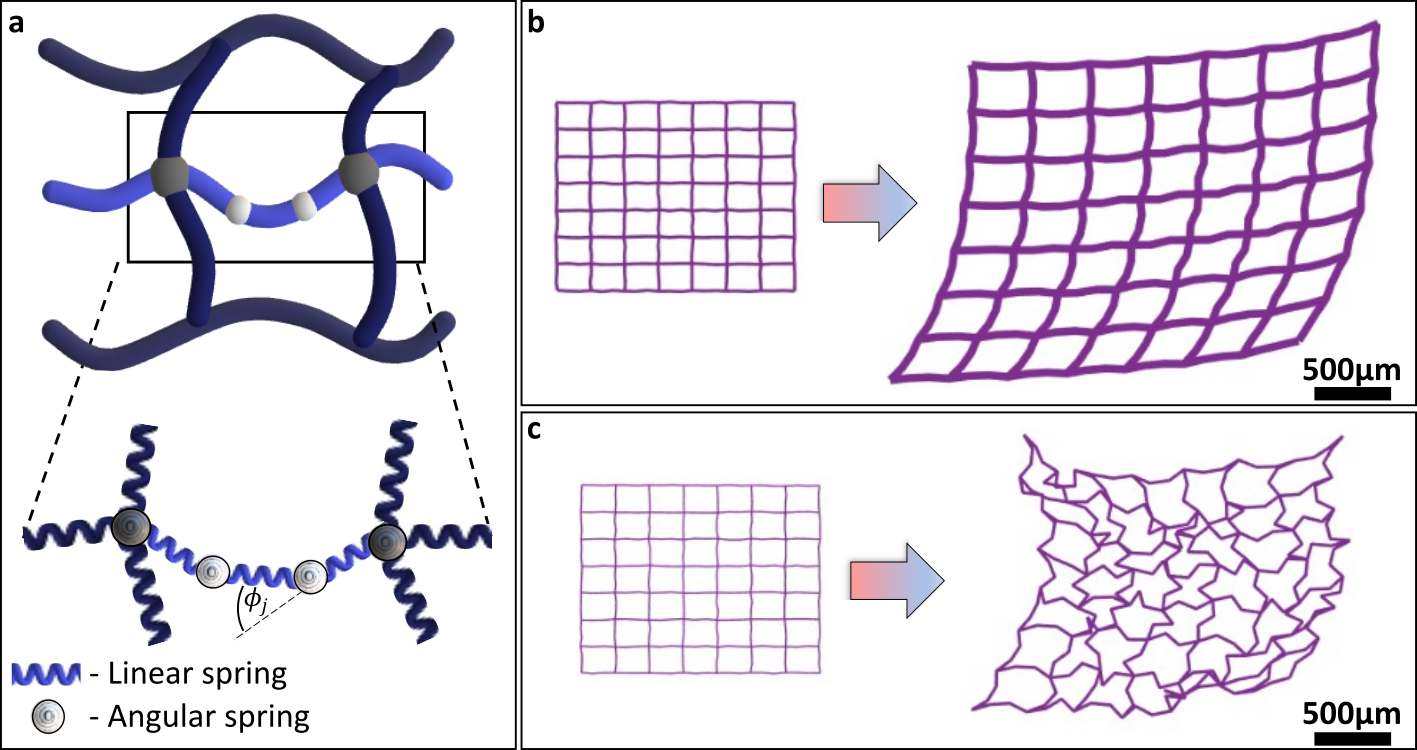}
    \caption{\textbf{(a)} A schematic illustration of the springs-network model. Each edge is connected at the nodes on both sides with three other edges. Each edge is constructed of a chain of linear springs, and every two springs are interconnected through an angular spring. \textbf{(b)} Simulation of a network at the initial state, with fiber diameter of 10 $\mu m$ (left), and at the final state (right) after the numerical swelling process, OP=$0.989$ (SD $0.011$); and \textbf{(c)} simulation of network at the initial state, with fiber diameter of 3 $\mu m$ (left), and at the final state (right) after the numerical swelling process, OP=$0.80$ (SD $0.22$).}
    \label{fig:sim}
\end{figure}

To analyze the effect of the mechanical parameters of the fibers, and in particular the role of their radius, on determining the morphing behavior of the network, we simulated the swelling process of the network using an empirical spring-based numerical model, which based on the structural mechanics and the elastic forces that are applied on each fiber according to Eq. \ref{energy}.
Within the model, the Cartesian network is constructed of edges that are connected at the nodes, where each node acts as an angular spring and connects four edges. Each edge is represented by a chain of linear springs that are connected by angular springs (Fig. \ref{fig:sim}a). For simplicity, we assume that all the springs throughout the network are similar, i.e. their spring constants and rest lengths are similar. The number of angular springs along each edge, and hence the number of linear springs that represent an edge, was set to the smallest number of springs that can reproduce a buckling length of 150$\mu m$ observed experimentally for the swollen messy systems. 

To model the swelling of the fibers and the resulting shape-morphing of the network, the swelling process was divided into a sequence of consecutive steps starting at the as-spun configuration and ending in the fully swollen state of the fibers. The change in the degree of swelling at each step was reflected in three parameters: the rest length of the linear springs, the radius of the fibers, and the Young modulus of the polymer. The last two parameters were used to calculate the stiffness and bending moduli, according to Eq. \ref{stiff}.  

Under the assumption that the evolution of the shape morphing follows a minimum energy path, the forces on all angular springs are calculated after each change of the parameters. Based on the energy of the system, described on Eq. \ref{energy}, the  force on the $j^{th}$ angular spring is given by:
\begin{equation}
    \bm{F_j}=k\sum_{n}(l_0-l_n)\bm{\hat{l_n}}+B\Big(\frac{\phi_j}{l_n+l_{n+1}}\Big)^2\bm{\hat{\alpha}},
\end{equation}
where $n$ account for  the number of linear springs connected at the angular spring, $l_0$ is the rest length of the linear spring, $l_n$ is the actual length of the $n^{th}$ spring, $\bm{\hat{l_n}}$ is the associated unit vector, $\phi_j$ is the angle between the springs (shown on {Fig. \ref{fig:sim}}a.), and $\bm{\hat{\alpha}}$ is the direction of the bending force. 

To prevent spatial overlap of the fibers and to account for excluded volume effects, a steep repulsive force is given by: 
\begin{equation}
    \bm{F_{overlap}(d_{j,k})}= \begin{cases}
    -e^{-2(d_{j,k}-2r)}\bm{\hat{d_{j,k}}} & d_{j,k}>2r, \\
    -1 & d_{j,k}\leq 2r,
    \end{cases}
\end{equation}
where $d_{j,k}$ is the distance between the $j^{th}$ and $k^{th}$ angular springs, and $\bm{\hat{d_{jk}}}$ is the associated unit vector. When the distance between the angular springs is smaller than the diameter of the fiber, then a maximal repulsion force is applied. As the distance increases, the repulsion force diminishes quickly.

The total force on an angular spring is $\bm{F}=\bm{F_j}+\bm{F_{overlap}(d_{j,k})}$.
Each angular spring is iteratively moved according to the calculated force. Once the amplitude of all the forces decreases below a predefined limit, the rest length, radius, and the Young modulus are altered again, and the process is repeated until the fibers reach their final swollen length.

The structural parameters of the model were taken from the average experimental results. According to the buckling length, each edge is divided into three springs. Initially, the length of the edges is set to $200 \mu m$ (the rest length of the comprising linear springs is set to $66.67 \mu m$). At the swollen state, the length of the edges is set to $350\mu m$ (the rest length of the linear springs is set to $116.67 \mu m$). At each iteration, the rest length, and radius of the spring are slightly increased until reaching the final state longitudinal swelling ratio of 1.75 and radial swelling ratio of 2.5, as was obtained experimentally. . The Young modulus, on the other hand, is slightly decreased in each iteration, ranging from $4000kPa$\supercite{Backes2018} for the unswollen as-spun configuration to $29.4kPa$ in the swollen state as calculated according to the Flory-Rehner theory (see Supplementary Information). Using the model described above, the behavior of different networks was simulated for diameters ranging from $3 \mu m$ to $30 \mu m$. To introduce inhomogeneity in the system, at the initiation, the positions of the angular springs are randomly slightly shifted from the perfect Cartesian lattice positions. 

Two typical examples of simulated shape morphing of networks with different fiber diameters are shown in Figs. \ref{fig:sim}, b and c.  The swelling simulation for a network composed of fibers with an unswollen diameter of $10 \mu m$ is shown in Fig. \ref{fig:sim}b. Despite a slight shearing, the overall shape of the network was preserved throughout the swelling process. The simulated swelling of a network with an unswollen fiber diameter of $3 \mu m$ is shown in Fig. \ref{fig:sim}c. Despite the similar structural features and Young modulus between the two systems, in this case the swollen state exhibited significant buckling, and the Cartesian shape of the network was not retained.

To compare the correlation between the simulated and the experimental data, the same order parameter was also applied on the simulated networks, and the values for different simulated networks with diameters ranging from 3 to 30 $\mu m$ are depicted on top of the experimental values in Fig. \ref{fig:morph}c (solid red line). The simulated results and the SD fit well to the experimental values, indicating that the change in morphing behavior can indeed be attributed to the change in mechanical properties, which is mainly affected by the radius of the filaments. In both cases, below a critical diameter of $10 \mu m $ the networks exhibit buckling-induced morphing and a relatively low order parameter, and a large standard deviation. Above that critical diameter, the order parameter is approximately 1 with a much smaller standard deviation, indicating a highly ordered morphology. 

The model implies that for the given Young modulus of the swollen polymer, the difference in behavior stems directly from the change in the radius of the fibers. Furthermore, the model confirms that the mechanical loads that are formed in the shape-morphing process are in the range of the critical load for fiber buckling. Hence small changes in the diameter of the filaments may result in a significant difference in shape-morphing behavior. Although these results were extracted for a specific polymeric system, the Young modulus of the polymer used in this study is within the range of many other natural and synthetic polymeric hydrogel systems. Therefore, we expect that a similar transition in morphing behaviors may occur in networks of microscale filaments constructed from other active and responsive materials. 

The ability to tune the shape-morphing behavior of a hierarchical structure of mesoscale building blocks without changing the polymer itself, but only by changing the diameter of the filaments and the network architecture, provides a novel, versatile, and scalable strategy for the design and development of novel functional responsive materials with high spatial morphing resolutions and with unique morphing abilities. Such systems pave the way for soft synthetic micro-muscles and micro-actuators with high spatial resolutions. Owing to the small dimensions of the filaments, they also hold significant potential for morphing optic devices and tunable separators. Lastly, such systems demonstrate that hierarchical structures of mesoscale elements may induce unexpected mechanical behaviors and that the dimensions of these elements must be accounted for, whether when dealing with synthetic hydrogel fibers or when examining biological systems such as microtubules and actin-myosin networks.

\noindent \textbf{Methods}\\
\noindent \textbf{Materials.}
N-isopropylacrylamide (NIPAAM) 99$\%$, Glycidyl methacrylate (GMA), and 2,2'-Azobis (2-methylpropionitrile) (AIBN) were purchased from Alfa Aesar. Toluene (anhydrous, 99.8$\%$), dimethylformamide (DMF), chloroform, dichloromethane (DCM), hexane, and tetraethylenepentamine (TEPA), poly[(m-pheneylenevinylene)-alt-(2,5-dihexyloxy-p-phenyleneviylene)] were purchased from Sigma-Aldrich. Diethyl ether and was purchased from BioLab. All the materials were used as bought, without further purification.\\

\noindent\textbf{Copolymer synthesis.}
The copolymer poly (N-isopropylacrylamide-co-glycidyl methacrylate) (PNcG) was synthesized using a radical polymerization through a modification of the procedure described by Xiaowei \textit{et al.} \supercite{Jiang2007}.  In a typical synthesis, NIPAAm (6.0 g, 0.053 mol), GMA (0.145 mL, 1.060 mmol), and AIBN (50 mg, 0.304 mmol) were dissolved in 40 mL of toluene, under a purge of Argon, and the monomers were allowed to polymerized at 70$^\circ$C for 2 hours in 100 mL three-neck flask. The solution was cooled down to room temperature and the solvent was evaporated using a rotary evaporator. The copolymer was then redissolved in DCM and was precipitated by slowly pouring 1 L of diethyl ether into the solution. The copolymer was filtered from the solution, dissolved in DCM and hexane (1:1 v/v). Then the solvent was evaporated using a rotary evaporator to yield a white powder. The product was dried under vacuum overnight. \\

\noindent\textbf{Fabrication of the networks.}
Fabrication of nano- to microscale fibers of PNIPPAm and its copolymers was demonstrated before using electrospinning \supercite{Rockwood2008,Okuzaki2009,Liu2016,Jiang2015}. However, obtaining high enough jet stability for accurate jet-writing using electrospinning is not trivial and hence dry-spinning was chosen as a jetting method instead to improve the stability \supercite{Imura2014,Brown2011,Moroni2006,Jordahl2018,Moon2021}. In this approach, the copolymer and the multi-arm amine crosslinker tetraethylenepentamine (TEPA) were first mixed and dissolved in the appropriate solvent mixture to form a viscous solution. In a typical jetting solution PNIPAAm copolymer of 0.5-0.7 g mL$^{-1}$ was prepared by dissolving the PNcG in a mixture of chloroform and DMF (1:1 v/v). 1$\%$-10$\%$ of TEPA was added just before the jetting process. At this range, the crosslinking is sufficient for preventing the dissolving of the copolymer in water, and yet there is a significant swelling of the hydrogel. 
Usually, trace amounts of the fluorescent polymeric dye were added to the solution to allow examination of the fibers in fluorescence microscopy. Next, the solution was dispensed via a metallic 23-gauge needle at a constant flow rate of 0.020 mL$\cdot$h$^{-1}$. Once a droplet was formed at the end of the capillary, it was mechanically pulled toward the rotating collector. The increase in surface area and the pulling process results in the evaporation of the solvents and in solidification into a thin fiber of mesoscale diameter. Once attached to the collector, the motion of the collector derives further pulling of the fiber. For the thin fibers (in average, fiber diameter $<$ 5 $\mu m$), the tip-to-ground distance was 3 cm, and the rotating speed of the drum was 16 - 32 mm/sec. The linear motion stage velocity was 1.15 mm$\cdot$sec$^{-1}$.
For the thick fibers (fiber diameter $>$ 5 $\mu m$), the tip-to-ground distance was 1 cm, and rotating speed of the drum was 1 - 5 mm/sec. The linear motion stage velocity was 0.2 mm$\cdot$sec$^{-1}$.
For a drum with a radius \(R_{d}\) and an angular velocity of \(\omega_{d}\), positioned on a stage moving at a velocity \(V_{s}\) in perpendicular to the drum rotation, the distance between the parallel fibers is given by \(V_{s}/(R_{d} \cdot \omega_{d}) \), and hence, for a given rotation speed, the distance between the fibers was controlled by the velocity of the moving stage.

To construct the networks, the fibers were collected on a plastic frame that was attached to a rotating drum positioned on top of a moving stage. At the first stage, fibers were pulled by the rotating drum along the direction of the rotation and perpendicular to the motion of the moving stage. The diameter of the fibers is determined by the drum velocity and by the polymer concentration of the jetting solution. The distance between the fibers is dictated by the rotation speed of the drum and the velocity of the moving stage. 

Once the fibers were jetted along the entire plastic frame, the frame was detached from the rotating drum and re-attached after rotating it by the angle. For constructing Cartesian networks the frame was rotated by an angle of 90$^{\circ}$, and the jet writing was of parallel fibers was repeated to obtain the final network structure. We define the crossing points of the fibers as the nodes of the network and segment of the fiber connecting two nodes as an edge. The crosslinking of the copolymer was performed post-fabrication of the network by baking it at 70$^{\circ}$C overnight. 
The experimental setup contained a syringe pump (New Era), a linear motion stage (ILS-200LM, Newport), a 8-axis univesal contoller (XPS-D8, Newport), and a rotating drum collector. \\

\noindent\textbf{Instrumentation.}
$^1$H-Nuclear magnetic resonance ($^1$H-NMR) spectra was recorded on Bruker  Avance III 400MHz spectrometers. The copolymer was dissolved in deuterated chloroform as a solvent. The chemical shifts are reported in ppm and referenced to the solvent. \\
Gel permeation chromatography (GPC) measurements were performed on Viscotek GPCmax by Malvern.\\
Scanning electron microscope (SEM) was performed using a Quanta 200FEG environmental SEM in a high vacuum, WD 10mm, 12.5-20kV. \\
All images and videos were taken by Olympus, IX73 microscope equipped with a heating glass slide (LCI, CU-301).

\section*{References}
\begin{enumerate}
\item Liu, Y., Genzer, J. \& Dickey, M. D. "2D or not 2D": Shape-programming polymer sheets. \textit{Progress in Polymer Science} \textbf{52}, 79–106 (2016).
\item Gracias, D. H. Stimuli responsive self-folding using thin polymer films. \textit{Current Opinion in Chemical Engineering} \textbf{2}, 112–119 (2013).
\item White, T. J. \& Broer, D. J. \textit{Nature Materials} \textbf{14}, 1087–1098 (2015).
\item Luo, Y., Lin, X., Birui, C. \& Xiaoyue, W. Cell-laden four-dimensional bioprinting using near-infrared-triggered shape-morphing alginate / polydopamine bioinks. \textit{Biofabrication} \textbf{11}, 045019 (2019).
\item Hines, L., Petersen, K., Lum, G. Z. \& Sitti, M. Soft Actuators for Small-Scale Robotics. \textit{Advanced Materials} \textbf{29}, 1603483 (2017).
\item Park, N. \& Kim, J. Hydrogel-Based Artificial Muscles: Overview and Recent Progress. \textit{Advanced Intelligent Systems} \textbf{2}, 1900135 (2020).
\item Sun, Y. L., Dong, W. F., Yang, R. Z., Meng, X., Zhang, L., Chen, Q. D. \& Sun, H. B. Dynamically tunable protein microlenses. \textit{Angewandte Chemie - International Edition} \textbf{51}, 1558–1562 (2012).
\item Dong, L., Agarwal, A. K., Beebe, D. J. \& Jiang, H. Adaptive liquid microlenses activated by stimuli-responsive hydrogels. \textit{Nature} \textbf{442}, 551–554 (2006).
\item Dong, L. \& Jiang, H. Autonomous microfluidics with stimuli-responsive hydrogels. \textit{Soft Matter} \textbf{3}, 1223–1230 (2007).
\item Ter Schiphorst, J., Saez, J., Diamond, D., Benito-Lopez, F. \& Schenning, A. P. Lightresponsive polymers for microfluidic applications. \textit{Lab on a Chip} \textbf{18}, 699–709 (2018).
\item Klein, Y., Efrati, E. \& Sharon, E. \textit{Science} \textbf{315}, 1116–1120 (2007).
\item Stoychev, G., Puretskiy, N. \& Ionov, L. Self-folding all-polymer thermoresponsive microcapsules. \textit{Soft Matter} \textit{7}, 3277–3279 (2011).
\item Zakharchenko, S., Puretskiy, N., Stoychev, G., Stamm, M. \& Ionov, L. Temperature controlled encapsulation and release using partially biodegradable thermo-magneto-sensitive self-rolling tubes \textit{Soft Matter} \textit{6}, 2633–2636 (2010). 
\item Wu, Z. L., Moshe, M., Greener, J., Therien-aubin, H., Nie, Z., Sharon, E. \& Kumacheva, E. Three-dimensional shape transformations of hydrogel sheets induced by small-scale modulation of internal stresses. \textit{Nature Communications} \textbf{4}, 1586 (2013).
\item Bae, J., Bende, N. P., Evans, A. A., Na, J. H., Santangelo, C. D. \& Hayward, R. C. Programmable and reversible assembly of soft capillary multipoles. \textit{Materials Horizons} \textbf{4}, 228–235 (2017).
\item Zhang, X., Pint, C. L., Lee, M. H., Schubert, B. E., Jamshidi, A., Takei, K., Ko, H., Gillies, A., Bardhan, R., Urban, J. J., Wu, M., Fearing, R. \& Javey, A. Optically- and thermallyresponsive programmable materials based on carbon nanotube-hydrogel polymer composites. \textit{Nano Letters} \textbf{11}, 3239–3244 (2011).
\item Gladman, A. S., Elisabetta A. Matsumoto, Ralph G. Nuzzo, Mahadevan, L. \& Lewis, J. A. Biomimetic 4D Printing. \textit{Nature Materials} \textbf{15}, 413–419 (2016).
\item 18. Boley, J.W., Van Rees,W. M., Lissandrello, C., Horenstein, M. N., Truby, R. L., Kotikian, A., Lewis, J. A. \& Mahadevan, L. Shape-shifting structured lattices via multimaterial 4D printing. \textit{Proceedings of the National Academy of Sciences of the United States of America} \textbf{116}, 20856–20862 (2019).
\item Chang, J., He, J., Mao, M., Zhou, W., Lei, Q., Li, X., Li, D., Chua, C. K. \& Zhao, X. Advanced material strategies for next-generation additive manufacturing. \textit{Materials} \textbf{11} (2018).
\item Fan, X., Chung, J. Y., Lim, Y. X., Li, Z. \& Loh, X. J. Review of Adaptive Programmable Materials and Their Bioapplications. \textit{ACS Applied Materials and Interfaces} \textbf{8}, 33351–33370 (2016).
\item Liu, Q.,Wang,W., Reynolds, M. F., Cao, M. C., Miskin, M. Z., Arias, T. A., Muller, D. A., McEuen, P. L. \& Cohen, I. Micrometer-sized electrically programmable shape-memory actuators for low-power microrobotics. \textit{Science Robotics} \textbf{6}, eabe6663 (2021).
\item Kaehr, B. \& Shear, J. B. Multiphoton fabrication of chemically responsive protein hydrogels for microactuation. \textit{Proceedings of the National Academy of Sciences of the United States of America} \textbf{105}, 8850–8854 (2008).
\item Lee, M. R., Phang, I. Y., Cui, Y., Lee, Y. H. \& Ling, X. Y. Shape-shifting 3D protein microstructures with programmable directionality via quantitative nanoscale stiffness modulation. \textit{Small} \textbf{11}, 740–748 (2015). 
\item Wei, S., Liu, J., Zhao, Y., Zhang, T., Zheng, M., Jin, F., Dong, X., Xing, J. \& Duan, X. Protein-Based 3D Microstructures with Controllable Morphology and pH-Responsive Properties. \textit{ACS Applied Materials and Interfaces} \textbf{9}, 42247–42257 (2017).
\item Leong, T. G., Zarafshar, A. M. \& Gracias, D. H. Three-dimensional fabrication at small size scales. \textit{Small} \textbf{6}, 792–806 (2010).
\item  Kempaiah, R. \& Nie, Z. From nature to synthetic systems: Shape transformation in soft materials. \textit{Journal of Materials Chemistry B} \textbf{2}, 2357–2368 (2014).
\item  Erb, R. M., Sander, J. S., Grisch, R. \& Studart, A. R. Self-shaping composites with programmable bioinspired microstructures. \textit{Nature Communications} \textbf{4}, 1–8 (2013).
\item  Kelby, T. S., Wang, M. \& Huck, W. T. Controlled folding of 2D Au-polymer brush composites into 3D microstructures. \textit{Advanced Functional Materials} \textbf{21}, 652–657 (2011).
29. Ge, Q., Qi, H. J. \& Dunn, M. L. Active materials by four-dimension printing. \textit{Applied Physics Letters} \textbf{103}, 131901 (2013).
\item  Meng, L., Klinkajon, W., K-hasuwan, P. R., Harkin, S., Supaphol, P. \& Wnek, G. E. Electrospun crosslinked poly(acrylic acid) fiber constructs: Towards a synthetic model of the cortical layer of nerve. \textit{Polymer International} \textbf{64}, 42–48 (2015).
\item  Jiang, S., Liu, F., Lerch, A., Ionov, L. \& Agarwal, S. Unusual and Superfast Temperature-Triggered Actuators. \textit{Advanced Materials} \textbf{27}, 4865–4870 (2015).
\item  Jalani, G., Jung, C.W., Lee, J. S.\& Lim, D.W. Fabrication and characterization of anisotropic nanofiber scaffolds for advanced drug delivery systems. \textit{International Journal of Nanomedicine} \textbf{9}, 33–49 (2014).
\item  Moon, S., Jones, M. S., Seo, E., Lee, J., Lahann, L., Jordahl, J. H., Lee, K. J. \& Lahann, J. 3D jet writing of mechanically actuated tandem scaffolds. \textit{Science Advances} \textbf{7}, 1–9 (2021).
\item  Apsite, I., Stoychev, G., Zhang, W., Jehnichen, D., Xie, J. \& Ionov, L. Porous Stimuli- Responsive Self-Folding Electrospun Mats for 4D Biofabrication. \textit{Biomacromolecules} \textbf{18}, 3178–3184 (2017).
\item  Zhao, Q., Wang, J., Cui, H., Chen, H., Wang, Y. \& Du, X. Programmed Shape-Morphing Scaffolds Enabling Facile 3D Endothelialization. \textit{Advanced Functional Materials} \textbf{28}, 1801027 (2018).
\item Meza, L. R., Das, S. \& Greer, J. R. Strong, lightweight, and recoverable three-dimensional ceramic nanolattices. \textit{Science} \textbf{345}, 1322–1326 (2014).
\item Moroni, L., DeWijn, J. R. \& Van Blitterswijk, C. A. 3D fiber-deposited scaffolds for tissue engineering: Influence of pores geometry and architecture on dynamic mechanical properties. \textit{Biomaterials} \textbf{27}, 974–985 (2006).
\item Chang, D., Liu, J., Fang, B., Xu, Z., Li, Z., Liu, Y., Brassart, L., Guo, F., Gao,W. \& Gao, C. Reversible fusion and fission of graphene oxide-based fibers. \textit{Science} \textbf{372}, 614–617 (2021).
\item Mueller, J., Lewis, J. A. \& Bertoldi, K. Architected Multimaterial Lattices with Thermally Programmable Mechanical Response. \textit{Advanced Functional Materials}, 2105128 (2021).
\item Halperin, A., Kröger, M. \& Winnik, F. M. Poly(N-isopropylacrylamide) Phase Diagrams: Fifty Years of Research. \textit{Angewandte Chemie} \textbf{54}, 15342–15367 (2015).
\item Schild, H. G. Poly(N-isopropylacrylamide): experiment, theory and application. \textit{Progress in Polymer Science} \textbf{17}, 163–249 (1992).
\item Audoly, B. \& Pomeau, Y.\textit{ Elasticity and Geometry from Hair Curls to the Non-linear Response of Shells chap. 1-3} (2010).
\item Backes, S.\& von Klitzing, R. Nanomechanics and nanorheology of microgels at interfaces. \textit{Polymers} \textbf{10}, 978 (2018).
\item Jiang, X., Xiong, D., An,Y., Zheng, P., Zhang,W.\& Shi, L. Thermoresponsive Hydrogel of Poly(glycidyl methacrylate-co-N-isopropylacrylamide) as a Nanoreactor of Gold Nanoparticles. \textit{Journal of Polymer Science:Part A:Polymer Chemistry} \textbf{45}, 2812–2819 (2007).
\item Rockwood, D.N., Chase, D. B., Akins, R. E. \& Rabolt, J. F. Characterization of electrospun poly(N-isopropyl acrylamide) fibers. \textit{Polymer} \textbf{49}, 4025–4032 (2008).
\item Okuzaki, H., Kobayashi, K. \& Yan, H. Non-woven fabric of poly(N-isopropylacrylamide) nanofibers fabricated by electrospinning. \textit{Synthetic Metals} \textbf{159}, 2273–2276 (2009).
\item Liu, L., Jiang, S., Sun,Y. \& Agarwal, S. Giving Direction to Motion and Surface with Ultra- Fast Speed Using Oriented Hydrogel Fibers. \textit{Advanced Functional Materials} \textbf{26}, 1021-1027 (2016).
\item Imura, Y., Hogan, R. M. \& Jaffe, M. A\textit{dvances in Filament Yarn Spinning of Textiles and Polymers chap. 3} (2014).
\item Brown, T. D., Dalton, P. D. \& Hutmacher, D. W. Direct writing by way of melt electrospinning. \textit{Advanced Materials} \textbf{23}, 5651–5657 (2011).
\item Jordahl, J. H., Solorio, L., Sun, H., Ramcharan, S., Teeple, C. B., Haley, H. R., Lee, K. J., Eyster, T.W., Luker, G. D., Krebsbach, P. H. \& Lahann, J. 3D JetWriting: Functional Microtissues Based on Tessellated Scaffold Architectures. \textit{Advanced Materials} \textbf{30}, 1707196
(2018).
\end{enumerate}

\noindent \textbf{Acknowledgments}\\
We would like to thank Prof. Michael Urbakh and Prof. Oded Hod for the insightful discussions A. Sitt acknowledges the generous support from the Azrieli Foundation. S. Ziv Sharabani and N. Edelsteien-Pardo acknowledge the generous support of The Shulamit
Aloni Scholarship for Advancing Women in Exact Science and Engineering, provided by The Ministry of Science $\&$ Technology, Israel. The authors acknowledge the Chaoul Center for Nanoscale Systems of Tel Aviv University for the use of instruments and staff assistance, , and the Mechanical Workshop for Research and Development, School of Chemistry, Tel Aviv University, for their help in constructing the fabrication devices.
\\
\textbf{Author contributions}: S. Ziv Sharabani and A.Sitt. conceived of and designed the project. S. Ziv Sharabani and N. Edelstein-Pardo synthesized the copolymer. S. Ziv Sharabani and M. Molco carried out SEM measurements. S. Ziv Sharabani, M. Morami, N. Bachar Schwartz, and A. Sivan prepared the samples, and carried out optical microscopy measurements. S. Ziv Sharabani analyzed the data, and carried out numerical model.  N. Edelstein-Pardo carried out $^{1}$H-NMR and GPC measurements. E. Flaxer conceived and constructed the fabrication devices. A. Sitt supervised the research.

\end{document}